\documentclass[onecolumn]{IEEEtran}
\usepackage{amsmath}
\usepackage{subcaption}
\usepackage{graphicx}
\usepackage{txfonts}
\usepackage{hyperref}

\usepackage[T1]{fontenc}
\usepackage[utf8]{inputenc}
\usepackage{authblk}

\title{Parameter Estimation for Intermediate-Mass Binary Black Holes through Gravitational Waves Observed by DECIGO}

\author[1]{MengFei Sun}
\author[1]{Jin Li \thanks{Corresponding author: cqujinli1983@cqu.edu.cn}}

\affil[1]{College of Physics, Chongqing University, Chongqing 401331, China}
\begin{document}
\maketitle % 这里放置标题页

\begin{abstract}
With the anticipated launch of space-based gravitational wave detectors, including LISA, TaiJi, TianQin, and DECIGO, expected around 2030, the detection of gravitational waves generated by intermediate-mass black hole binaries (IMBBHs) becomes a tangible prospect. 
However, due to the detector's reception of a substantial amount of non-Gaussian, non-stationary data, employing traditional Bayesian inference methods for parameter estimation would result in significant resource demands and limitations in the waveform template library. 
Therefore, in this paper, we simulated foreground noise induced by stellar-origin binary black holes (SOBBHs), which is non-Gaussian and non-stationary, and we explore the use of Gaussian process regression (GPR) and deep learning for parameter estimation of Intermediate Mass Binary Black Holes (IMBBHs) in the presence of such non-Gaussian, non-stationary background noise.
By comparing these results from deep learning and GPR, we demonstrate that deep learning can offer improved precision in parameter estimation compared to traditional GPR. 
Furthermore, compared to GPR, deep learning can provide posterior distributions of the sample parameters faster.
\end{abstract}

\section{Introduction}
Intermediate-Mass Black Holes (IMBHs) have garnered significant attention in the field of astrophysics. 
They occupy a mass range that falls between Stellar-Mass Black Holes and Supermassive Black Holes, typically ranging from $10^{2} M_{\odot}$ to $10^{5} M_{\odot}$.
While we understand that IMBHs have masses ranging from $10^{2} M_{\odot}$ to $10^{5} M_{\odot}$, there remain numerous unresolved questions regarding their origins, formation, and evolutionary processes. 
Investigating intermediate-mass black holes not only contributes to unveiling the evolutionary pathways of Supermassive Black Holes in the universe but also aids in gaining a deeper understanding of gravitational wave astronomy, stellar cluster dynamics, and the formation and evolution of cosmic structures.
Up to the present, the most likely scenario for the formation of IMBHs involves the collapse and merger process of super massive stars. 
These super massive stars form in the early stages of the lifetime of a young star cluster. While it's unlikely for a single star to form a black hole exceeding 100 solar masses, within star clusters, a multitude of main-sequence stars may undergo a series of recurrent collisions and mergers, leading to the formation of IMBHs exceeding 100 solar masses \cite{1,2,3,4,5,6}.
IMBHs formed through collisions and mergers, undergo accretion of additional material from within the star cluster over time, leading to a further increase in their mass \cite{7,8,9}.
In 2006, Gurkan et al. systematically investigated the process of forming two massive stars in dense star clusters \cite{10}. 
Their findings suggest that the formation of two supermassive stars is possible in sufficiently dense central regions of star clusters, further indicating the potential for the formation of Intermediate-Mass Binary Black Holes (IMBBHs) in the dense cores of clusters.
If IMBBHs exist at the center of a galaxy, the IMBHs will gradually approach each other through dynamical interactions, entering an inspiraling phase, and eventually merging into a larger black hole. During this process, the IMBBHs system will continuously emit gravitational waves, which are expected to be detectable by the gravitational wave detectors.

With the anticipated launch of space-based gravitational wave detectors, including LISA, TaiJi, TianQin, and DECIGO, expected around 2030. 
LISA, TaiJi, and TianQin are primarily focused on detecting Supermassive Black Holes and the stochastic gravitational wave background (SMBHs) within the mass range of $10^{5} M_{\odot}$ to $10^{8} M_{\odot}$ in the millihertz frequency band. 
DECIGO's detection objectives include  both Intermediate-Mass Black Holes (IMBHs) and the stochastic gravitational wave background in the Deci-Hertz frequency band \cite{34}.
In the study, we choose the Deci-Hertz Interferometer Gravitational Wave Observatory (DECIGO) as our observational tool. DECIGO is a future mission plan of the Japanese space program designed to detect gravitational waves within the frequency range of $f\sim 0.1 - 10$ Hz \cite{11,12,13}. 
It is worth noting that gravitational waves from white dwarf/white dwarf binary systems are truncated at $f = 0.2$ Hz \cite{14}, thus rendering DECIGO's high-frequency regime immune to such interfering noise.
However, according to the Stellar-origin Binary Black Hole (SOBBH) population model \cite{15,16,17,18,19,20,21}, a substantial amount of gravitational waves originating from Stellar-origin binary black holes may superpose, forming the non-Gaussian and non-stationary gravitational wave background noise. 
In this study, we use 56747 SOBBH events, each characterized by different mass and luminosity distance values, as an integral component of our background noise model. This component of the background noise exhibits non-Gaussian and non-stationary features, adding practical significance and complexity to the task of recovering source parameters for IMBBHs.

The method for analyzing gravitational wave in the presence of noise backgrounds typically is Bayesian approach \cite{75}. In the field of ground-based Gravitational Wave (GW) detection, the approach has been widely employed for parameter estimation of Stellar-Mass Binary Black Holes  \cite{36,37,38}. However, this method requiring a lot of computing power. As the LIGO/Virgo collaborations find more gravitational wave (GW) events, the cost of using this approach becomes more expensive. When using it in space-based GW detection, the challenges will become bigger. 
In recent years, deep learning methods, particularly Variational Autoencoders and Convolutional Neural Networks, have made significant strides in the rapid and precise extraction of parameter information from gravitational wave data generated by detectors in gravitational wave parameter estimation. Research has demonstrated the feasibility of deep learning in the gravitational wave parameter estimation process \cite{22,23,24,25,26,27,28,33}.
In this study, we propose three neural network models, CNN-Multivariate Gaussian Distribution (CMGD), Dense-Multivariate Gaussian Distribution (DMGD), and LSTM-GRU-Multivariate Gaussian Distribution (LRMGD).
Through these three neural networks and the traditional Gaussian Process Regression (GPR) method \cite{50,51}, we conducted parameter estimation on simulated data. These three neural networks and the GPR method provided posterior distributions for five parameters of IMBBHs: $(m1, m2, chirp mass, z, D_{L})$. By comparing the deep learning methods with the traditional Gaussian Process Regression (GPR) method, our study demonstrates that the neural networks we propose can more precisely constrain the parameters of IMBBHs and can provide posterior distributions of the sample parameters faster.

The article is structured as follows: Section 2 presents our method for simulating gravitational wave waveforms and background noise. Section 3 discusses the Gaussian Process Regression (GPR) method and provides a detailed description of the GPR model we selected. Section 4 introduces the neural networks we designed, including the model architecture and hyperparameters. In Section 5, we show the results obtained by the GPR model and the three neural networks in extracting the parameter distributions of IMBBHs from non-stationary, non-Gaussian background noise. Finally, our main conclusions and final remarks are presented in Section 6.

\section{Data simulation}

\subsection{Distribution of IMBBHs and SOBBH}

For the distribution of IMBBHs and SOBBH, we adopt the following form of \cite{41}:
\begin{equation}
	\begin{aligned}
		\rho(z) \sim \frac{4 \pi d_{C}^{2}(z) R(z)}{H(z)(1+z)},
	\end{aligned}
\end{equation}
where the co-moving distance is $d_{C}(z) = \int_{0}^{z} 1 / H\left(z^{\prime}\right) d z^{\prime}$, and the evolution of the inflation rate with time is quantified as \cite{42,43,44}:
\begin{equation}
	\begin{aligned}
		R(z)=\left\{\begin{array}{cc}
			1+2 z, & z \leq 1 \\
			\frac{3}{4}(5-z), & 1<z<5, \\
			0, & z \geq 5
		\end{array}\right.
	\end{aligned}
\end{equation}
the above distribution is normalized as: 
\begin{equation}
	\begin{aligned}
		\rho(z)=\frac{4 \pi a d_{C}^{2}(z) R(z)}{H(z)(1+z)},
	\end{aligned}
\end{equation}
where a is normalization factor.

\subsection{Simulation of Gravitational Wave Signal}

We employed pycbc, which includes TaylorT2 waveform templates, to simulate gravitational wave signals, with TaylorT2 primarily used to describe the inspiral gravitational waveform of binary systems \cite{39,40}. Based on the distribution in Eq.~3, we simulated 10,000 IMBBHs with varying masses. Given that DECIGO's sensitive frequency range spans from 0.1 Hz to 10 Hz, we selected a mass range for intermediate-mass binary black holes between $(10^{3}M_{\odot},10^{4}M_{\odot})$. The gravitational wave frequencies from binary black hole mergers in this mass range fall within DECIGO's sensitive range, as illustrated in Figure 1.

\begin{figure}[htbp]
	\centering
	\includegraphics[width=0.5\textwidth]{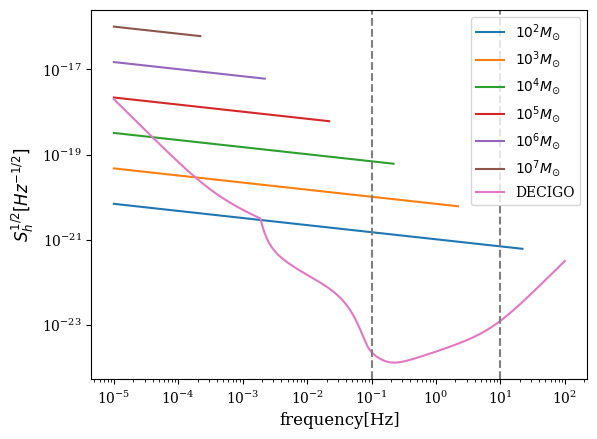}    
	\caption{
The pink line represents the noise power spectral density (PSD) of DECIGO, while the blue and yellow lines depict the frequency-domain amplitudes of gravitational waves from binary black holes with masses of $10^{3}M_{\odot}$ and $10^{4}M_{\odot}$, respectively.
    }
\end{figure}

We set the parameters for IMBBHs as follows: $m_1$, $m_2$ are drawn from a Uniform distribution between $10^{3}M_{\odot}$ and $10^{4}M_{\odot}$, the sampling rate is 10Hz, the Cut-off frequency $f_{low}$ is 0.1Hz, and the luminosity distance to the source is:
\begin{equation}
	\begin{aligned}
		D_{L}=\frac{1+z}{H_{0}} \int_{0}^{z} \frac{d 	z^{\prime}}{\left[\Omega_{M}\left(1+z^{\prime}\right)^{3}+\Omega_{\Lambda}\right]^{1 / 2}}.
	\end{aligned}
\end{equation}

The redshifts range from z $\in [0, 5]$, and the redshift distribution follows Eq.~3. Our simulation is conducted within the framework of the flat $\Lambda \mathrm{CDM}$ model, with $\Omega_{m}=0.31$, $\Omega_{\Lambda}=0.69$, and $H_{0}=67.74$\cite{45}. Finally, we randomly select 50 seconds of data from each of the simulated 10000 signals, resulting in a data length of 500 for our desired signals.

\subsection{Simulation of Noise}

The  noise power spectral density of DECIGO is given by \cite{46,47}: 
\begin{equation}
	\begin{aligned}
	S_{h}^{\text {DECIGO }}(f) & =7.05 \times 10^{-48}\left[1+\left(\frac{f}{f_{p}}\right)^{2}\right]+4.8 \times 10^{-51}\left(\frac{f}{1 H z}\right)^{-4} \\
	& \times\frac{1}{1+\left(\frac{f}{f_{p}}\right)^{2}}+5.53 \times 10^{-52}\left(\frac{f}{1 H z}\right)^{-4} Hz^{-1},
	\end{aligned}
\end{equation}
where $f_{p}=7.36 Hz$. The three terms correspond to short noise, radiation pressure noise, and acceleration noise, respectively.

Based on the one-sided noise power spectral density (PSD), we could get the time-domain noise signal from the one-sided PSD, in this paper, we utilized the Python function pycbc.noise.gaussian.noise\_from\_psd (which takes a PSD as input and returns colored Gaussian noise) to get $n_{1}(t)$.

Additionally, we have considered the potential presence of a substantial gravitational wave noise background originating from Stellar-Origin Binary Black Holes (SOBBHs) within the redshift range of 0 to 5\cite{49}. The formation rate of SOBBHs is referenced from\cite{21}, and in this study, we have adopted a formation rate of $R_{\mathrm{BHB}} = 27 \mathrm{Gpc}^{-3} \mathrm{yr}^{-1}$. The comoving volume within the redshift range of 0 to 5 is based on the Planck 2018 results\cite{48}, with $V_{\mathrm{comoving}} = 2101.76 \mathrm{Gpc}^{3}$. Our observation time is set to one year.
Therefore, we simulated waveforms from 56747 SOBBHs and superimposed them, with the redshift distribution of these 56747 SOBBHs following Eq.~3. To maintain consistency with the model used for the signals, we employed TaylorT2 waveforms as templates for SOBBHs. The parameters we set for SOBBHs were: m1, m2=Uniform$(3M_{\odot},41M_{\odot})$\cite{21}, sampling rate is 10Hz, the Cut-off frequency $f_{low}$=0.1Hz, and the luminosity distance to the source follows Eq.~4. Subsequently, we randomly selected 50 seconds of data from each simulated SOBBHs waveforms, resulting in a data length of 500 for superposition. By superimposing these 56747 SOBBHs waveforms, we formed our $n_{2}(t)$, the non-Gaussian and non-stationary gravitational wave background noise from SOBBHs.
Finally, we add the obtained detector noise $n_{1}(t)$ and the stochastic gravitational wave background from SOBBHs $n_{2}(t)$ together and filter them within the frequency range of DECIGO's sensitivity, which is 0.1Hz to 10Hz. This process yields our ultimate non-stationary, non-Gaussian background noise:
\begin{equation}
	\begin{aligned}
n=[n_{1}(t)+n_{2}(t)]*F_{filter},
	\end{aligned}
\end{equation}
where $F_{filter}$ represents the filter that retains signals within the frequency range of 0.1Hz to 10Hz.

\subsection{Construction of Datasets}

The input data for both the neural network and GPR is as follows:
\begin{equation}
	\begin{aligned}
    s=h(t)+n,
	\end{aligned}
\end{equation}
where
\begin{equation}
	\begin{aligned}
h(t)=F_{+}(\bar{\theta}_{\mathrm{S}}, \bar{\phi}_{\mathrm{S}}, \bar{\theta}_{\mathrm{L}}, \bar{\phi}_{\mathrm{L}}) h_{+}(t)+F_{\times}(\bar{\theta}_{\mathrm{S}}, \bar{\phi}_{\mathrm{S}}, \bar{\theta}_{\mathrm{L}}, \bar{\phi}_{\mathrm{L}}) h_{\times}(t),
	\end{aligned}
\end{equation}
here $F_{+,x}$ are the antenna pattern functions. $(\bar{\theta}_{\mathrm{S}}, \bar{\phi}_{\mathrm{S}})$ is the source ecliptic Latitude and longitude, $\left(\bar{\theta}_{\mathrm{L}}, \bar{\phi}_{\mathrm{L}}\right)$ is the
direction of the orbital angular momentum in the solar barycentric frame. In the framework of the detector, the antenna pattern functions can be written as follows:
\begin{subequations}
\begin{align}
F_{+}\left(\theta_{\mathrm{S}}, \phi_{\mathrm{S}}, \psi_{\mathrm{S}}\right)=\frac{1}{2}\left(1+\cos ^{2} \theta_{\mathrm{S}}\right) \cos \left(2 \phi_{\mathrm{S}}\right) \cos \left(2 \psi_{\mathrm{S}}\right)-\cos \left(\theta_{\mathrm{S}}\right) \sin \left(2 \phi_{\mathrm{S}}\right) \sin \left(2 \psi_{\mathrm{S}}\right), \\
F_{\times}\left(\theta_{\mathrm{S}}, \phi_{\mathrm{S}}, \psi_{\mathrm{S}}\right)=\frac{1}{2}\left(1+\cos ^{2} \theta_{\mathrm{S}}\right) \cos \left(2 \phi_{\mathrm{S}}\right) \sin \left(2 \psi_{\mathrm{S}}\right)+\cos \left(\theta_{\mathrm{S}}\right) \sin \left(2 \phi_{\mathrm{S}}\right) \cos \left(2 \psi_{\mathrm{S}}\right),
\end{align}
\end{subequations}
here, the two angles $(\theta_{\mathrm{S}}, \phi_{\mathrm{S}})$ represent the direction of the source in the detector frame and  $\psi_{\mathrm{S}}$ is the polarization angle given as:

\begin{subequations}
\begin{align}
\theta_{\mathrm{S}}(t) & =\cos^{-1} \left (  \frac{1}{2} \cos \bar{\theta}_{\mathrm{S}}-\frac{\sqrt{3}}{2} \sin \bar{\theta}_{\mathrm{S}} \cos \left[\bar{\phi}(t)-\bar{\phi}_{\mathrm{S}}\right] \right ),\\
\phi_{\mathrm{S}}(t) & =\frac{\pi}{12}+\tan ^{-1}\left(\frac{\sqrt{3} \cos \bar{\theta}_{\mathrm{S}}+\sin \bar{\theta}_{\mathrm{S}} \cos \left[\bar{\phi}(t)-\bar{\phi}_{\mathrm{S}}\right]}{2 \sin \bar{\theta}_{\mathrm{S}} \sin \left[\bar{\phi}(t)-\bar{\phi}_{\mathrm{S}}\right]}\right),\\
\psi_{\mathrm{S}}&=\tan^{-1} \left ( \frac{a}{b}  \right ) ,\\
a &= \frac{1}{2} \cos \bar{\theta}_{\mathrm{L}}-\frac{\sqrt{3}}{2} \sin \bar{\theta}_{\mathrm{L}} \cos \left[\bar{\phi}(t)-\bar{\phi}_{\mathrm{L}}\right] \notag \\
&- \cos \bar{\theta}_{\mathrm{L}} \cos^{2}  \bar{\theta}_{\mathrm{S}}-\sin \bar{\theta}_{\mathrm{L}} \sin \bar{\theta}_{\mathrm{S}} \cos \left(\bar{\phi}_{\mathrm{L}}-\bar{\phi}_{\mathrm{S}}\right),   \\
b &= \frac{1}{2} \sin \bar{\theta}_{\mathrm{L}} \sin \bar{\theta}_{\mathrm{S}} \sin \left(\bar{\phi}_{\mathrm{L}}-\bar{\phi}_{\mathrm{S}}\right)  \notag \\
 &-\frac{\sqrt{3}}{2} \cos \bar{\phi}(t)\left(\cos \bar{\theta}_{\mathrm{L}} \sin \bar{\theta}_{\mathrm{S}} \sin \bar{\phi}_{\mathrm{S}}-\cos \bar{\theta}_{\mathrm{S}} \sin \bar{\theta}_{\mathrm{L}} \sin \bar{\phi}_{\mathrm{L}}\right) \notag \\
 &-\frac{\sqrt{3}}{2} \sin \bar{\phi}(t)\left(\cos \bar{\theta}_{\mathrm{S}} \sin \bar{\theta}_{\mathrm{L}} \cos \bar{\phi}_{\mathrm{L}}-\cos \bar{\theta}_{\mathrm{L}} \sin \bar{\theta}_{\mathrm{S}} \cos \bar{\phi}_{\mathrm{S}}\right),\\
 \bar{\theta}(t)&=\pi / 2, \\
 \bar{\phi}(t)&=2 \pi t / T+c_{0},
\end{align}
\end{subequations}
Eq.~9 provides the expression for the response function in the detector frame, while Eq.~10 presents the transformation formula between the detector frame and the solar system frame,from $(\theta_{\mathrm{S}}, \phi_{\mathrm{S}}, \psi_{\mathrm{S}})$ to $(\bar{\theta}_{\mathrm{S}}, \bar{\phi}_{\mathrm{S}}, \bar{\theta}_{\mathrm{L}}, \bar{\phi}_{\mathrm{L}})$. Actually,DECIGO is moving around the sun with the period T = 1yr and characterized by
the position angles $\bar{\theta}$ and $\bar{\phi}$.

In Section 2.2, we simulated 10,000 IMBBHs; however, we retained only 6,621 signals with a signal-to-noise ratio (SNR) $\in [0,100]$, as shown in Figures 2 and 3. Among these 6,621 data points, 80\% were allocated for the training set, denoted as x\_train, while the remaining 20\% constituted the testing set, denoted as x\_test. Correspondingly, the labels, denoted as y\_train and y\_test, represent the parameter space, which includes $(m1, m2, chirp mass, z, D_{L})$, required by both the neural network and GPR for training and testing.

\begin{figure}[htbp]
	\centering
	\includegraphics[width=0.5\textwidth]{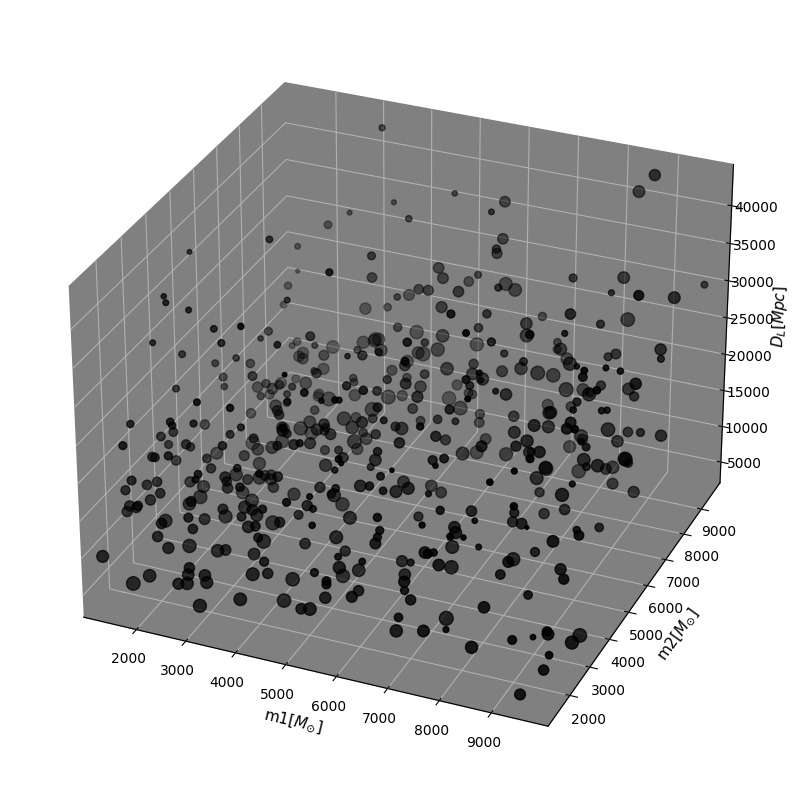}    
	\caption{The x and y coordinates represent the masses of the two IMBHs, while the z coordinate corresponds to the luminosity distance. Additionally, the size of the points in the figure represents the relative signal-to-noise ratio (SNR) of the sources within our simulated noisy background.
    }
\end{figure}

\begin{figure}[htbp]
	\centering
	\includegraphics[width=0.5\textwidth]{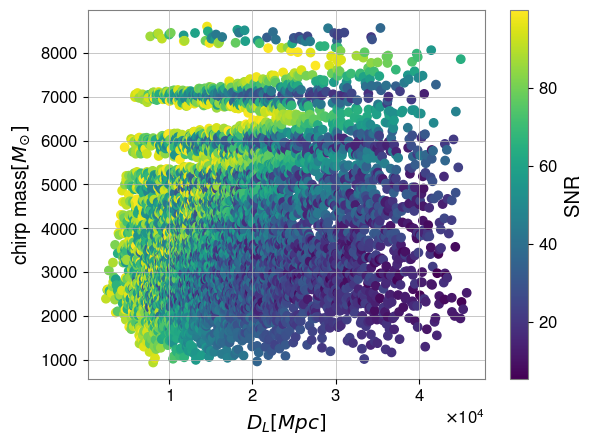}    
	\caption{The x-coordinate represents the luminosity distance, the y-coordinate represents the chirp mass of the IMBBHs system, and the color bar indicates the SNR of each signal.
    }
\end{figure}

\section{Gaussian Process Regression}
\subsection{GPR Model Introduction}
Gaussian Process Regression (GPR) is a powerful non-parametric probabilistic model\cite{50,52,53} that utilizes Gaussian processes (GPs) to model random functions. GPR is employed for regression problems, where the goal is to predict a continuous output variable. Unlike traditional regression methods, it does not require specifying the shape or parameters of the function in advance and can provide not only predictions but also prediction uncertainties. This capability makes GPR suitable for parameter estimation in gravitational wave data. The detailed mathematical principles of GPR can be found in reference \cite{50}. Below, we provide a brief overview of the GPR process. 
Let  $\mathcal{X}$  be any set. A Gaussian process (GP) is a collection of random variables indexed by  $\mathcal{X}$  such that if  $\left\{X_{1}, \ldots, X_{n}\right\} \subset \mathcal{X}$  is any finite subset, the marginal density  $p\left(X_{1}=x_{1}, \ldots, X_{n}=x_{n}\right)$  is multivariate Gaussian. Any Gaussian distribution is completely specified by its first and second central moments (mean and covariance), and GP's are no exception. We can specify a GP completely in terms of its mean function  $\mu: \mathcal{X} \rightarrow \mathbb{R}$  and covariance function  k: $\mathcal{X} \times \mathcal{X} \rightarrow \mathbb{R}$ . Most of the expressive power of GP's is encapsulated in the choice of covariance function. 
We assume that the target function is a random function that can be represented using a Gaussian process, which is a collection of infinitely many random variables such that the joint distribution of any finite subset is a multivariate Gaussian distribution. It can be expressed using the following mathematical formula:

\begin{equation}
	\begin{aligned}
f(x) \sim G P\left(m(x), k\left(x, x^{\prime}\right)\right),
	\end{aligned}
\end{equation}
in this formula, $f(x)$ represents the target function, $x$ is the input variable, $m(x)$ is the mean function, and $k(x, x^\prime)$ is the kernel function. This equation signifies that for any set of input variables, the distribution of $f(x)$ is a Gaussian distribution with its mean determined by $m(x)$ and its covariance determined by $k(x, x^\prime)$. In GPR, we utilize observed data to infer the parameters of the mean function and kernel function, thereby obtaining the posterior distribution of the parameters.

In GPR, we assume that we have a set of observed data: $\left\{\left(x_{i}, y_{i}\right)\right\}_{i=1}^{n}$ where $x{i}$ represents inputs and $y_{i}$ represents the corresponding outputs. We aim to predict the output $y_{}$ for a given input $x_{}$. We can predict both the mean and the covariance matrix.

The mean of $y_{*}$ is given by the mean of the posterior predictive distribution:
\begin{equation}
	\begin{aligned}
\mu_{*}=\mathbb{E}\left[y_{*} \mid x_{*}, \mathbf{X} \mathbf{y}\right],
 	\end{aligned}
\end{equation}
where $\mathbf{X}$ represents the input set of observed data, and $\mathbf{y}$ is the output set of observed data, according to the properties of Gaussian processes, the predictive mean $\mu_{}$ is simply the value of the mean function $m\left(x_{}\right)$ at the point $x_{*}$:
$\mu_{*}=m\left(x_{*}\right)$.

the covariance of $y_{*}$ is provided by the covariance of the posterior predictive distribution:
\begin{equation}
	\begin{aligned}
\operatorname{Var}\left[y_{*} \mid x_{*}, \mathbf{X}, \mathbf{y}\right]=k\left(x_{*}, x_{*}\right)-\mathbf{k}_{*}^{T}\left(\mathbf{K}+\sigma_{n}^{2} \mathbf{I}\right)^{-1} \mathbf{k}_{*},
 	\end{aligned}
\end{equation}
where $k(x_{}, x_{})$ represents the self-covariance of $x_{}$, $\mathbf{k}_{}$ is the covariance vector between $x_{*}$ and all observed data points, $\mathbf{K}$ is the covariance matrix among observed data points, $\sigma_{n}^{2}$ is the noise variance, and $\mathbf{I}$ is the identity matrix.

\subsection{The Selection and Hyperparameters of Kernel Functions}
In this paper, we have selected four kernel functions that have shown promising predictive performance for this task: the Radial basis function kernel, the Rational Quadratic kernel, the Matern kernel, and the Dot-Product kernel\cite{50}:

Radial basis function kernel is given by:
\begin{equation}
	\begin{aligned}
k\left(x_{i}, x_{j}\right)=\exp \left(-\frac{d\left(x_{i}, x_{j}\right)^{2}}{2 l^{2}}\right),
 	\end{aligned}
\end{equation}
where l is the length scale of the kernel and  $d(\cdot, \cdot)$is the Euclidean distance.

Rational Quadratic kernel is given by:
\begin{equation}
	\begin{aligned}
k\left(x_{i}, x_{j}\right)=\left(1+\frac{d\left(x_{i}, x_{j}\right)^{2}}{2 \alpha l^{2}}\right)^{-\alpha},
 	\end{aligned}
\end{equation}
where $\alpha$ is the scale mixture parameter,l is the length scale of the kernel and $d(\cdot, \cdot)$ is the Euclidean distance. 

Matern kernel is given by:
\begin{equation}
	\begin{aligned}
k\left(x_{i}, x_{j}\right)=\frac{1}{\Gamma(\nu) 2^{\nu-1}}\left(\frac{\sqrt{2 \nu}}{l} d\left(x_{i}, x_{j}\right)\right)^{\nu} K_{\nu}\left(\frac{\sqrt{2 \nu}}{l} d\left(x_{i}, x_{j}\right)\right),
 	\end{aligned}
\end{equation}
where  $d(\cdot, \cdot)$ is the Euclidean distance, $K_{\nu}(\cdot)$  is a modified Bessel function and $\Gamma(\cdot)$ is the gamma function. 

Dot-Product kernel is given by:
\begin{equation}
	\begin{aligned}
k\left(x_{i}, x_{j}\right)=\sigma_{0}^{2}+x_{i} \cdot x_{j},
 	\end{aligned}
\end{equation}
where $\sigma_{0}$ controls the inhomogenity of the kernel.

We employed these four kernel functions to perform Gaussian process regression on our data, and the optimal hyperparameters for each of these kernel functions are presented in Table 1. The posterior distribution plots of our gravitational wave parameters obtained through these four kernel functions are illustrated in Figure 5.

\begin{table*}[htbp]
\caption{The optimal hyperparameters for the four kernel functions}
\begin{tabular}{lllllll}
\hline
   & ConstantKernel & WhiteKernel(noise\_level) & l(length\_scale) & $\alpha$ & $\nu$ & $\sigma_{0}$  \\ 
   \hline
Radial basis function kernel & $0.032^{2}$    & 0.0164                    & 369              &          &       &              \\ 
Rational Quadratic kernel    & $0.775^{2}$    & 0.0143                    & 961              & 0.00976  &       &              \\ 
Matern kernel                & $0.301^{2}$    & 0.0160                    & 405              &          & 6.5   &              \\ 
Dot-Product kernel           & $0.045^{2}$    & 0.0227                    &                  &          &       & 4.99         \\ \hline
\end{tabular}
\end{table*}

\section{Deep Learning}
\subsection{Neural Networks with Multivariate Gaussian Distribution}

In the past few years, traditional neural networks have excelled in many machine learning tasks, successfully applying point estimation methods to regression and classification problems. However, in certain critical application domains, particularly in scientific research such as gravitational wave parameter estimation, providing point estimates alone may not be sufficient to meet the requirements. Because point estimation cannot provide information about the uncertainty associated with parameter estimates, and in practical applications, understanding the degree of uncertainty in parameter estimation is often essential.
The neural network developed in this study employs a probabilistic modeling approach, utilizing multivariate Gaussian distributions to represent the probability distributions of parameters. This implies that the network's output goes beyond a mere point estimate and provides a probability distribution that accounts for parameter uncertainties. Specifically, the network can output both the mean and covariance matrix of the parameters, offering insights into the central tendencies and the range of parameter estimates.

Compared to traditional Gaussian Process Regression (GPR) methods, the neural network employed in this study exhibits significant distinctions and advantages in modeling multivariate Gaussian distributions of parameters. The neural network adopts the framework of deep learning, endowed with powerful representation learning capabilities that can adapt to complex nonlinear relationships. 
Neural networks can extract local features using Dense and CNN, while Gaussian Process Regression primarily considers global features, which can lead to overfitting or underfitting issues\cite{55}.

Additionally, in our neural network, we employ the Adam optimizer to continuously optimize model parameters with the aim of minimizing the negative log-likelihood loss between the distribution generated by the model and the distribution of observed data. This means that the core objective of training the neural network is to find a set of parameters that make the model's output as close as possible to the distribution of actual observed data.
In GPR, parameter search typically involves optimizing model hyperparameters, such as kernel function parameters. The goal of GPR is to find the best kernel function parameters to make the model fit the data as well as possible while considering uncertainty. Usually, to achieve this goal, methods like Maximum Marginal Likelihood Estimation (MMLE) or Bayesian optimization are employed to search for these hyperparameters. However, these methods often come with a high computational cost\cite{56}.

\subsection{Architecture and Hyperparameters of the Three Neural Networks}

We designed three neural networks: CNN-Multivariate Gaussian Distribution (CMGD), Dense-Multivariate Gaussian Distribution (DMGD), and LSTM-GRU-Multivariate Gaussian Distribution (LRMGD). These networks are specifically crafted to address the issue of uncertainty in gravitational wave parameter estimation. Each network employs a distinct architecture, making optimal use of key components of deep learning such as CNNs, dense (fully connected) layers, as well as LSTM and GRU units.
In the design of these networks, we placed a significant emphasis on feature extraction and data representation, which are pivotal challenges in gravitational wave parameter estimation. The initial layers of each network are dedicated to capturing salient features within the data, extracting them for subsequent utilization by the Multivariate Gaussian Distribution layers\cite{57}. This distribution layer is of paramount importance as it models the probability distribution of gravitational wave parameters, aiding in our understanding of parameter uncertainty.
To optimize the networks to approximate the most probable gravitational wave parameter distribution, we employed efficient optimizers and continually updated the network's parameters through iterative processes. This step is crucial in training the networks, gradually bringing them closer to the true parameter distribution and enhancing prediction accuracy.
We have meticulously documented the architecture and hyperparameter settings of these three neural networks, which are provided in Tables 2, 3, and 4, respectively. This documentation assists researchers in gaining a better understanding of the network configurations and how to replicate the experiments effectively.

These detailed records are essential for ensuring the reproducibility and robustness of our results in the field of gravitational wave parameter estimation.

\begin{table*}[htbp]
\caption{CMGD Network Architecture and Hyperparameter Configuration}
\begin{tabular}{llll}
\hline
Layer type                         & Parameters                                                                                                                      & Input     & Output                                                                                                                                                                                  \\ \hline
Dense                              & units=128                                                                                                                       & (500,1)   & (500,128)                                                                                                                                                                               \\ \hline
Conv1D                             & \begin{tabular}[c]{@{}l@{}}filters=128,kernel\_size=8,activation='relu',\\ MaxPooling1D(pool\_size=2),Dropout(0.1)\end{tabular} & (500,128) & (246,128)                                                                                                                                                                               \\ \hline
Conv1D                             & \begin{tabular}[c]{@{}l@{}}filters=64,kernel\_size=8,activation='relu',\\ Dropout(0.2)\end{tabular}                             & (246,128) & (239,64)                                                                                                                                                                                \\ \hline
Conv1D                             & \begin{tabular}[c]{@{}l@{}}filters=32,kernel\_size=8,activation='relu',\\ Dropout(0.1)\end{tabular}                             & (239,64)  & (232,32)                                                                                                                                                                                \\ \hline
Conv1D                             & \begin{tabular}[c]{@{}l@{}}filters=16,kernel\_size=8,activation='relu',\\ Dropout(0.2)\end{tabular}                             & (232,32)  & (225,16)                                                                                                                                                                                \\ \hline
Flatten                            &                                                                                                                                 & (225,16)  & (3600)                                                                                                                                                                                  \\ \hline
Dense                              & units=100                                                                                                                       & (3600)    & (100)                                                                                                                                                                                   \\ \hline
Multivariate Gaussian Distribution &   event\_shape=(5)           & (100)     & \begin{tabular}[c]{@{}l@{}}$N(\boldsymbol{\mu},\boldsymbol{\Sigma})$,\\ $\quad \boldsymbol{\mu} \in \mathbb{R}^{5}$, \\ $\quad \boldsymbol{\Sigma} \in \mathbb{R}^{5 x 5}$\end{tabular} \\ \hline
Loss                               & negative log-likelihood                                                                                                         &           &                                                                                                                                                                                         \\ 
Optimizer                          & Adam(learning rate=0.00008)                                                                                                     &           &                                                                                                                                                                                         \\ 
Epoch                              & 800                                                                                                                             &           &                                                                                                                                                                                         \\ 
Batch size                         & 16                                                                                                                              &           &                                                                                                                                                                                         \\ \hline
\end{tabular}
\end{table*}

\begin{table*}[htbp]
\caption{DMGD Network Architecture and Hyperparameter Configuration}
\begin{tabular}{llll}
\hline
Layer type                         & Parameters                                                        & Input & Output                                                                                                                                                                                  \\ \hline
Dense                              & units=128                                                         & (500) & (128)                                                                                                                                                                                   \\ \hline
Dense                              & \begin{tabular}[c]{@{}l@{}}units=300,\\ Dropout(0.1)\end{tabular} & (128) & (300)                                                                                                                                                                                   \\ \hline
Dense                              & \begin{tabular}[c]{@{}l@{}}units=400,\\ Dropout(0.1)\end{tabular} & (300) & (400)                                                                                                                                                                                   \\ \hline
Dense                              & \begin{tabular}[c]{@{}l@{}}units=500,\\ Dropout(0.2)\end{tabular} & (400) & (500)                                                                                                                                                                                   \\ \hline
Dense                              & \begin{tabular}[c]{@{}l@{}}units=600,\\ Dropout(0.1)\end{tabular} & (500) & (600)                                                                                                                                                                                   \\ \hline
Flatten                            &                                                                   & (600) & (600)                                                                                                                                                                                   \\ \hline
Dense                              & units=100                                                         & (600) & (100)                                                                                                                                                                                   \\ \hline
Multivariate Gaussian Distribution &   event\_shape=(5)               & (100) & \begin{tabular}[c]{@{}l@{}}$N(\boldsymbol{\mu}, \boldsymbol{\Sigma})$,\\ $\quad \boldsymbol{\mu} \in \mathbb{R}^{5}$,\\ $\quad \boldsymbol{\Sigma} \in \mathbb{R}^{5 x 5}$\end{tabular} \\ \hline
Loss                               & negative log-likelihood                                           &       &                                                                                                                                                                                         \\ 
Optimizer                          & Adam(learning rate=0.00008)                                       &       &                                                                                                                                                                                         \\ 
Epoch                              & 800                                                               &       &                                                                                                                                                                                         \\ 
Batch size                         & 16                                                                &       &                                                                                                                                                                                         \\  \hline
\end{tabular}
\end{table*}

\begin{table*}[htbp]
\caption{LRMGD Network Architecture and Hyperparameter Configuration}
\begin{tabular}{llll}
\hline
Layer type                         & Parameters                  & Input    & Output                                                                                                                                                                                  \\ \hline
LSTM                               & units=2000,Dropout(0.1)     & (1,500)  & (1,2000)                                                                                                                                                                                \\ \hline
GRU                                & units=1000,Dropout(0.2)     & (1,2000) & (1,1000)                                                                                                                                                                                \\ \hline
LSTM                               & units=1000,Dropout(0.2)     & (1,1000) & (1,1000)                                                                                                                                                                                \\ \hline
Multivariate Gaussian Distribution & event\_shape=(5)            & (1,1000) & \begin{tabular}[c]{@{}l@{}}$N(\boldsymbol{\mu}, \boldsymbol{\Sigma})$,\\ $\quad \boldsymbol{\mu} \in \mathbb{R}^{5}$,\\ $\quad \boldsymbol{\Sigma} \in \mathbb{R}^{5 x 5}$\end{tabular} \\ \hline
Loss                               & negative log-likelihood     &          &                                                                                                                                                                                         \\ 
Optimizer                          & Adam(learning rate=0.00008) &          &                                                                                                                                                                                         \\ 
Epoch                              & 800                         &          &                                                                                                                                                                                         \\ 
Batch size                         & 16                          &          &                                                                                                                                                                                         \\ \hline
\end{tabular}
\end{table*}

\section{Results and Analysis}

In our analysis, we employed pre-trained neural network models and Gaussian Process Regression (GPR) models to perform parameter sampling and analysis on a sample from the test dataset, as depicted in Figure 4. Initially, we utilized GPR models with four different kernel functions to perform 10,000 parameter samples on this sample. This process yielded posterior distributions for five parameters of the Intermediate Mass Black Hole (IMBH), namely, $(m1, m2, chirp mass, z, D_{L})$, and their joint distributions, as illustrated in Figure 5. Figure 6 presents the probability distributions of each parameter from the four GPR models, while Figure 7 showcases predictions of parameter means and standard deviations produced by these models.

From the analysis of Figures 5, 6, and 7, it is evident that in our study, the GPR model utilizing the Rational Quadratic kernel function outperforms in predicting the posterior distribution of sample parameters. Next, we will compare the best-performing result from the GPR models with our three designed neural networks. Figures 8, 9, and 10 present our comparative results.

In Tables 5 to 11, we provide the parameter covariance matrices predicted by the four GPR models and the three neural networks. Through calculations, we can conclude that the predictions of parameter means and variances by the three neural networks are more accurate than the best result among the four GPR models.

\begin{figure}[htbp]
	\centering
	\includegraphics[width=0.5\textwidth]{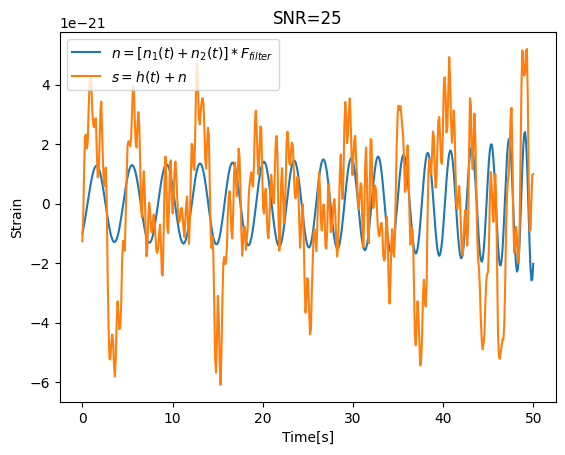}    
	\caption{
 The x axis represents time in seconds, and the y axis represents the gravitational wave strain. The duration of this sample is 50 seconds, with a sampling rate of 10Hz and a signal-to-noise ratio of 25. The blue curve represents the gravitational wave waveform during the inspiral phase of IMBBHs, while the orange curve represents the result of the superposition of noise and signal.
    }
\end{figure}

\begin{figure}[htbp]
	\centering
	\includegraphics[width=1\textwidth]{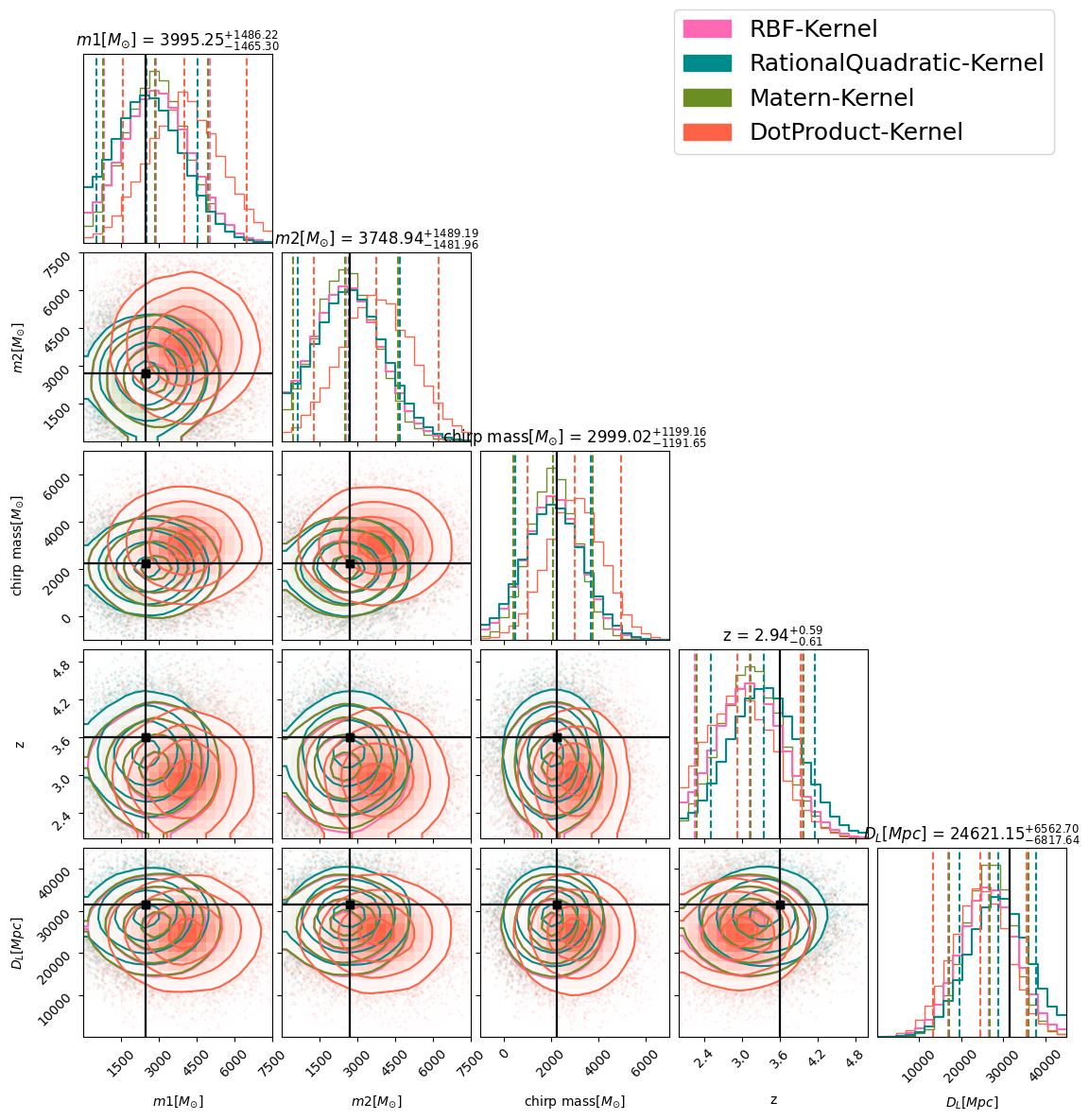}    
	\caption{Corner plot showing one and two-dimensional marginalized posterior distributions of the GW parameters $(m1, m2, chirp mass, z, D_{L})$ for one example test dataset. The contour boundaries enclose 68\%, 90\%, and 95\% probability regions. One-dimensional histograms of the posterior distribution for each parameter from the four GPR models are plotted along the diagonal. Vertical dashed lines in the one-dimensional plots represent the 5\%–95\% symmetric confidence bounds of the one-dimensional posteriors from the four GPR models. Black vertical and horizontal lines indicate the true parameter values of the simulated signal.
    }
\end{figure}

\begin{figure}[htbp]
	\centering
	\includegraphics[width=1\textwidth]{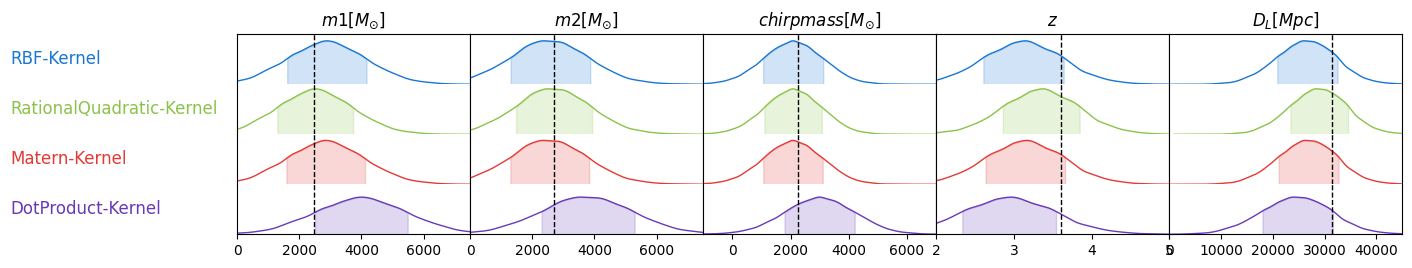}    
	\caption{
 The various colors represent GPR models corresponding to different kernel functions, displaying probability distribution plots for each parameter.
    }

\end{figure}

\begin{figure}[htbp]
	\centering
	\includegraphics[width=1\textwidth]{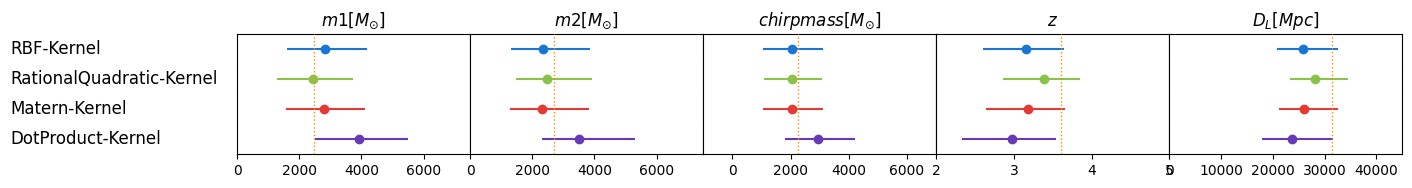}    
	\caption{
 The various colors represent the means and uncertainties of each parameter given by GPR models corresponding to different kernel functions.
    }

\end{figure}

\begin{figure}[htbp]
	\centering
	\includegraphics[width=1\textwidth]{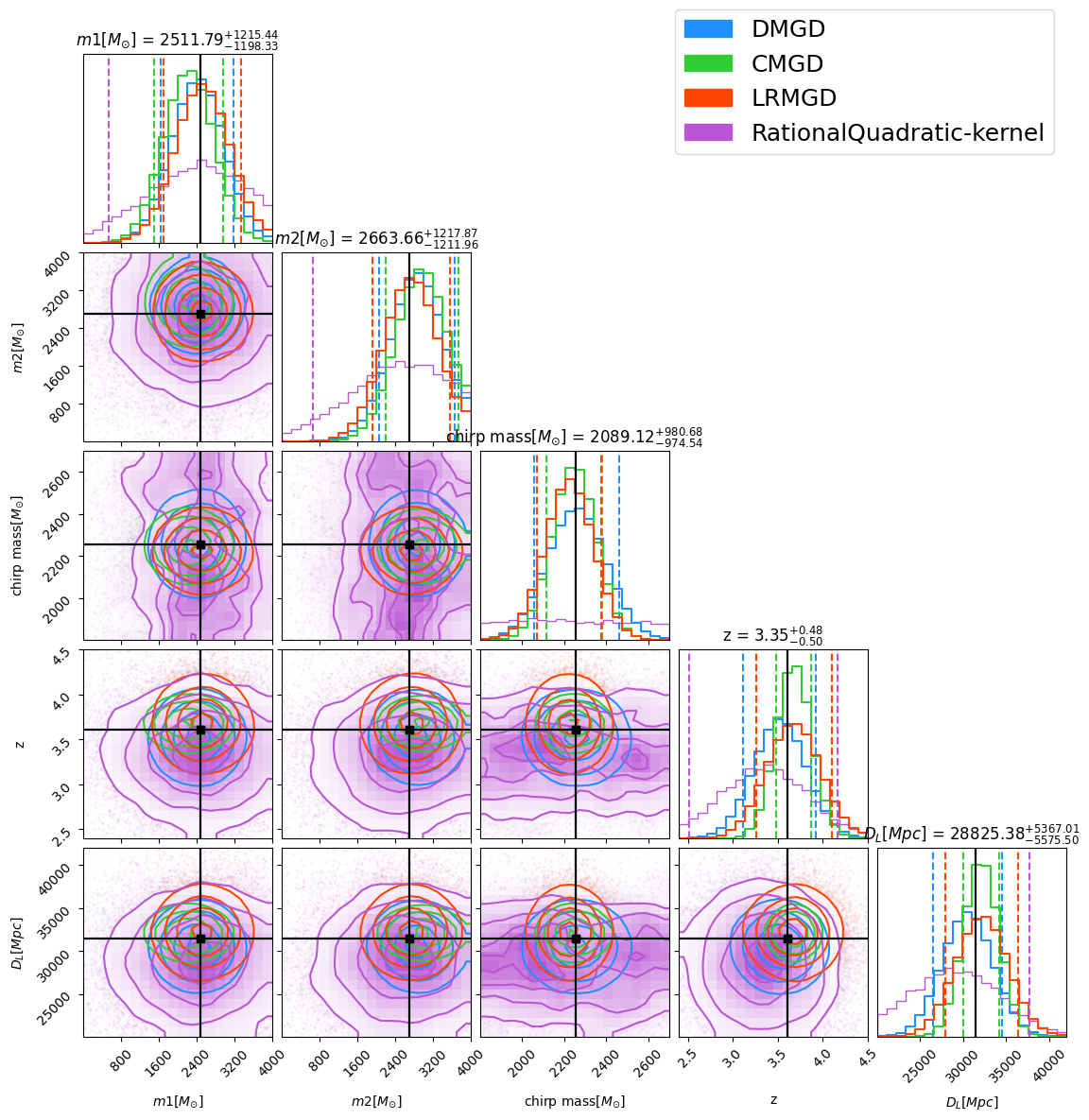}    
	\caption{
 Parameter Posterior Distributions: Corner plot displaying marginalized posterior distributions for GW parameters $(m1, m2, chirp mass, z, D_{L})$. Contours represent 68\%, 90\%, and 95\% probability intervals. One-dimensional histograms show parameter posteriors from RationalQuadratic-kernel-GPR and three neural networks. Vertical dashed lines denote 5\%-95\% symmetric confidence bounds, while black lines indicate true parameter values of the simulated signal.
    }

\end{figure}

\begin{figure}[htbp]
	\centering
	\includegraphics[width=1\textwidth]{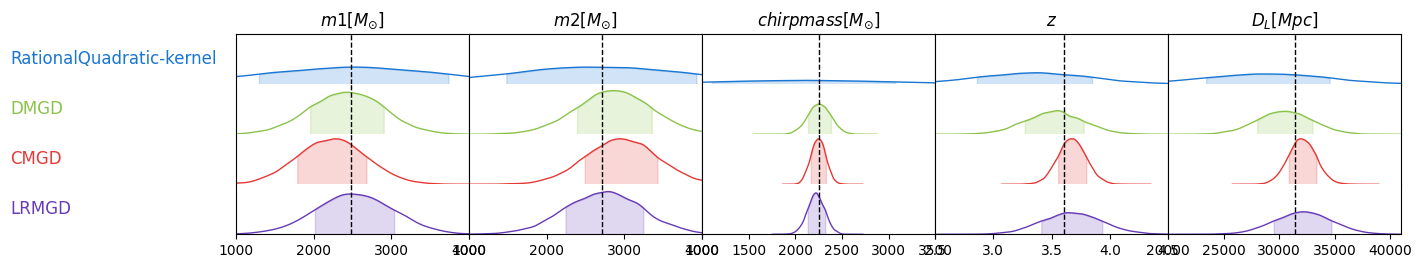}    
	\caption{
Different colors represent the probability distribution plots of each parameter provided by the RationalQuadratic-kernel-GPR and the three neural networks.
    }

\end{figure}

\begin{figure}[htbp]
	\centering
	\includegraphics[width=1\textwidth]{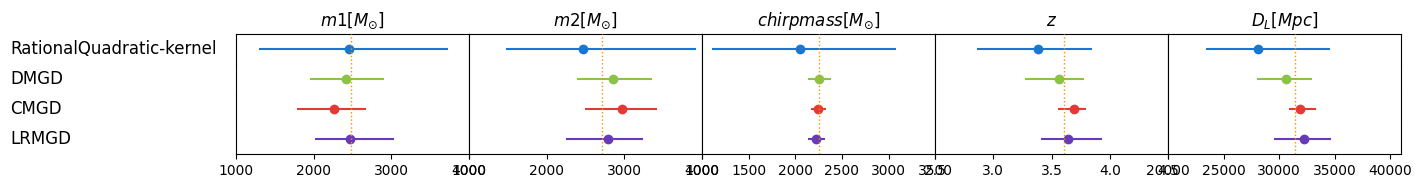}    
	\caption{
Different Colors Represent Means and Errors: Various colors indicate the means and errors of each parameter provided by the RationalQuadratic-kernel-GPR and the three neural networks.
    }

\end{figure}

\begin{table*}[htbp]
    \centering
    \caption{Parameter Covariance-RBF-Kernel}
    \begin{tabular}{c|ccccc}
         & $m1[M_{\odot}]$ & $m2[M_{\odot}]$ & $chirp mass[M_{\odot}]$ & z & $D_{L}[Mpc]$\\ 
        \hline
                $m1[M_{\odot}]$ & 1624945.88 & -12678.60 & 2830.88 &  0.92 & 87603.11 \\ 
                $m2[M_{\odot}]$ & -12678.60 & 1643089.66 & 8662.05 & 11.58 & -52079.26 \\ 
        $chirp mass[M_{\odot}]$ & 2830.88 & 8662.05 & 1060479.69 &  3.48 & -28923.57 \\ 
                            z &  0.92 & 11.58 &  3.48 &  0.27 & -37.05 \\ 
                   $D_{L}[Mpc]$ & 87603.11 & -52079.26 & -28923.57 & -37.05 & 33494468.80 \\ 
        \hline
    \end{tabular}
\end{table*}

\begin{table*}[htbp]
    \centering
    \caption{Parameter Covariance-RationalQuadratic-kernel}
    \begin{tabular}{c|ccccc}
         & $m1[M_{\odot}]$ & $m2[M_{\odot}]$ & $chirp mass[M_{\odot}]$ & z & $D_{L}[Mpc]$\\ 
        \hline
                m1$[M_{\odot}]$ & 1484101.57 & -11579.67 & 2585.51 &  0.84 & 80010.00 \\ 
                m2$[M_{\odot}]$ & -11579.67 & 1500672.71 & 7911.26 & 10.57 & -47565.22 \\ 
        $chirp mass[M_{\odot}]$ & 2585.51 & 7911.26 & 968561.22 &  3.18 & -26416.58 \\ 
                            z &  0.84 & 10.57 &  3.18 &  0.25 & -33.84 \\ 
                   $D_{L}[Mpc]$ & 80010.00 & -47565.22 & -26416.58 & -33.84 & 30591291.83 \\ 
        \hline
    \end{tabular}
\end{table*}

\begin{table*}[htbp]
    \centering
    \caption{Parameter Covariance-Matern-Kernel}
    \begin{tabular}{c|ccccc}
         & $m1[M_{\odot}]$ & $m2[M_{\odot}]$ & $chirp mass[M_{\odot}]$ & z & $D_{L}[Mpc]$\\ 
        \hline
                $m1[M_{\odot}]$ & 1592528.35 & -12425.66 & 2774.40 &  0.90 & 85855.44 \\ 
                $m2[M_{\odot}]$ & -12425.66 & 1610310.16 & 8489.24 & 11.35 & -51040.29 \\ 
        $chirp mass[M_{\odot}]$ & 2774.40 & 8489.24 & 1039323.21 &  3.41 & -28346.55 \\ 
                            z &  0.90 & 11.35 &  3.41 &  0.27 & -36.31 \\ 
                   $D_{L}[Mpc]$ & 85855.44 & -51040.29 & -28346.55 & -36.31 & 32826257.00 \\ 
        \hline
    \end{tabular}
\end{table*}

\begin{table*}[htbp]
    \centering
    \caption{Parameter Covariance-DotProduct-Kernel}
    \begin{tabular}{c|ccccc}
         & $m1[M_{\odot}]$ & $m2[M_{\odot}]$ & $chirp mass[M_{\odot}]$ & z & $D_{L}[Mpc]$\\ 
        \hline
                $m1[M_{\odot}]$ & 2219035.12 & -17313.97 & 3865.87 &  1.26 & 119631.30 \\ 
                $m2[M_{\odot}]$ & -17313.97 & 2243812.37 & 11828.94 & 15.81 & -71119.73 \\ 
        $chirp mass[M_{\odot}]$ & 3865.87 & 11828.94 & 1448196.95 &  4.75 & -39498.19 \\ 
                            z &  1.26 & 15.81 &  4.75 &  0.37 & -50.59 \\ 
                  $ D_{L}[Mpc]$ & 119631.30 & -71119.73 & -39498.19 & -50.59 & 45740232.60 \\ 
        \hline
    \end{tabular}
\end{table*}

\begin{table*}[htbp]
    \centering
    \caption{Parameter Covariance-DMGD}
    \begin{tabular}{c|ccccc}
         & $m1[M_{\odot}]$ & $m2[M_{\odot}]$ & $chirp mass[M_{\odot}]$ & z & $D_{L}[Mpc]$\\ 
        \hline
                $m1[M_{\odot}]$ & 222014.60 & 3678.93 & -112.14 &  0.46 & -6898.95 \\ 
                $m2[M_{\odot}]$ & 3678.93 & 233657.96 & -761.04 & -0.02 & -17094.84 \\ 
        $chirp mass[M_{\odot}]$ & -112.14 & -761.04 & 14960.10 & -0.49 & -2321.48 \\ 
                            z &  0.46 & -0.02 & -0.49 &  0.06 & -13.73 \\ 
                   $D_{L}[Mpc]$ & -6898.95 & -17094.84 & -2321.48 & -13.73 & 6073220.58 \\ 
        \hline
    \end{tabular}
\end{table*}

\begin{table*}[htbp]
    \centering
    \caption{Parameter Covariance-CMGD}
    \begin{tabular}{c|ccccc}
         & $m1[M_{\odot}]$ & $m2[M_{\odot}]$ & $chirp mass[M_{\odot}]$ & z & $D_{L}[Mpc]$\\ 
        \hline
                $m1[M_{\odot}]$ & 196714.47 & -607.29 & -337.45 &  0.57 & -4231.40 \\ 
                $m2[M_{\odot}]$ & -607.29 & 215663.77 & -313.78 &  0.51 & 2255.33 \\ 
        $chirp mass[M_{\odot}]$ & -337.45 & -313.78 & 6484.61 & -0.11 & 680.39 \\ 
                            z &  0.57 &  0.51 & -0.11 &  0.01 &  0.96 \\ 
                   $D_{L}[Mpc]$ & -4231.40 & 2255.33 & 680.39 &  0.96 & 1561850.10 \\ 
        \hline
    \end{tabular}
\end{table*}

\begin{table*}[htbp]
    \centering
    \caption{Parameter Covariance-LRMGD}
    \begin{tabular}{c|ccccc}
         & $m1[M_{\odot}]$ & $m2[M_{\odot}]$ & $chirp mass[M_{\odot}]$ & z & $D_{L}[Mpc]$\\ 
        \hline
                $m1[M_{\odot}]$ & 251561.65 & 2498.27 & -482.89 & -2.07 & 10218.74 \\ 
                $m2[M_{\odot}]$ & 2498.27 & 249951.28 & 789.54 & -0.11 & 12392.61 \\ 
        $chirp mass[M_{\odot}]$ & -482.89 & 789.54 & 8765.45 &  0.13 & 629.06 \\ 
                            z & -2.07 & -0.11 &  0.13 &  0.07 & -8.48 \\ 
                   $D_{L}[Mpc]$ & 10218.74 & 12392.61 & 629.06 & -8.48 & 6601945.03 \\ 
        \hline
    \end{tabular}
\end{table*}

\section{Summary and Discussion}

In this study, we simulated the non-stationary, non-Gaussian stochastic gravitational wave background noise from Stellar-origin Binary Black Holes. We extracted the parameter distribution of intermediate-mass black hole binaries (IMBBHs) from this simulated noise. We employed Gaussian Process Regression (GPR) with four different kernel functions and three neural networks (CMGD, DMGD, LRMGD) for regression to our simulated data.
By comparing the results in Figures 5 to 10 and Tables 5 to 11, it is evident that the neural networks constructed in this study outperform traditional Gaussian Process Regression in parameter estimation. Additionally, our approach enables more efficient parameter sampling.

It is worth emphasizing that the neural network developed in this study is not only applicable to providing posterior parameter distributions for gravitational wave data received by DECIGO (or other gravitational wave detectors such as LISA, TaiJi, TianQin). Finally, in future works, we can apply  deep learning methods to constrain the cosmological parameters associated with the selected scientific objectives encompassed by the DECIGO \cite{64,65,66,67,68,69,70,71,72,73}. This will open up a new window for gravitational-wave cosmology.

\section*{Acknowledgements}

Our code runs in the TensorFlow environment, and our Multivariate Gaussian Distribution layers are sourced from TensorFlow Probability \cite{57}. The Gaussian Process Regression code operates within the Scikit-Learn environment \cite{58}. Our figures are generated using matplotlib \cite{59} and ChainConsumer \cite{60}.

\bibliographystyle{IEEEtran}
\bibliography{bibliography}

\end{document}